\documentclass[aps,pra,reprint,superscriptaddress]{revtex4-2}

\usepackage{graphicx}
\usepackage{color}

\usepackage{xspace}

\newcommand{\MBNExplorer}
{\textsc{MBN Explorer}\xspace}

\newcommand{\MBNStudio}
{\textsc{MBN Studio}\xspace}

\begin{document}

\title{Irradiation driven molecular dynamics: A review}

\author{Alexey V. Verkhovtsev}
\altaffiliation{On leave from Ioffe Institute, Polytekhnicheskaya 26, 194021 St. Petersburg, Russia}
\affiliation{MBN Research Center, Altenh\"oferallee 3, 60438 Frankfurt am Main, Germany}

\author{Ilia A. Solov'yov}
\altaffiliation{On leave from Ioffe Institute, Polytekhnicheskaya 26, 194021 St. Petersburg, Russia}
\affiliation{Department of Physics, Carl von Ossietzky Universit\"at Oldenburg, Carl-von-Ossietzky-Str. 9-11, 26129 Oldenburg, Germany}

\author{Andrey V. Solov'yov}
\altaffiliation{On leave from Ioffe Institute, Polytekhnicheskaya 26, 194021 St. Petersburg, Russia}
\affiliation{MBN Research Center, Altenh\"oferallee 3, 60438 Frankfurt am Main, Germany}

\begin{abstract}
This paper reviews Irradiation Driven Molecular Dynamics (IDMD) -- a novel computational methodology for atomistic simulations of the irradiation driven transformations of complex molecular systems implemented in the \MBNExplorer software package. Within the IDMD framework various quantum processes occurring in irradiated systems are treated as random, fast and local transformations incorporated into the classical MD framework in a stochastic manner with the probabilities elaborated on the basis of quantum mechanics. Major transformations of irradiated molecular systems (such as topological changes, redistribution of atomic partial charges, alteration of interatomic interactions) and possible paths of their further reactive transformations can be simulated by means of MD with reactive force fields, in particular with the reactive CHARMM (rCHARMM) force field implemented in \MBNExplorer. This paper reviews the general concept of the IDMD methodology and the rCHARMM force field and provides several exemplary case studies illustrating the utilization of these methods.
\end{abstract}

\maketitle


\section{Introduction}
\label{sec:Introduction}

There are numerous examples of chemical transformations of complex molecular systems driven by irradiation.
Particular examples include (i) inactivation of living cells by ionizing radiation due to the induced complex DNA strand breaks \cite{schardt2010heavy, Surdutovich_2014_EPJD.68.353, solov2016nanoscale}; (ii) the formation and composition of cosmic ices in the interstellar medium and planetary atmospheres due to the interplay of the molecular surface adsorption and surface irradiation \cite{Tielens2013}; (iii) the formation of biologically relevant molecules under extreme conditions involving irradiation \cite{Horneck2002}, and many more.

\begin{sloppypar}
Irradiation driven chemistry (IDC) is nowadays utilized in modern nanotechnologies, such as focused electron beam induced deposition (FEBID) \cite{VanDorp2008, Utke_book_2012, Huth2012} and extreme ultraviolet (EUV) lithography \cite{Banqiu2007, hawryluk1988}. FEBID and EUV belong to the next generation of nanofabrication techniques allowing the controlled creation of complex three-dimensional nanostructures with nanometer resolution.
Fabrication of increasingly smaller structures has been the goal of the electronics industry for more than three decades and remains one of this industry's biggest challenges. Furthermore, IDC is a key element in nuclear waste decomposition technologies \cite{heidet2015} and medical radiotherapies \cite{schardt2010heavy, solov2016nanoscale, Linz2012_IonBeams}.
\end{sloppypar}

\begin{sloppypar}
IDC studies transformations of molecular systems induced by their irradiation with photon, neutron or charged-particle beams. IDC is also relevant for molecular systems exposed to external fields, mechanical stress, or plasma environment. A rigorous quantum-mechanical description of the irradiation-driven molecular processes e.g. within time-dependent density functional theory (TDDFT) is feasible but only for relatively small molecular systems containing, at most, a few hundred atoms \cite{Dinh2010, Maruhn2010, Jacquemin2009, Bochevarov2013}. This strong limitation makes TDDFT of limited use for the description of the IDC of complex molecular systems.
\end{sloppypar}

\begin{sloppypar}
Classical molecular dynamics (MD) represents an alternative theoretical framework for modeling complex molecular systems. For instance, the classical molecular mechanics approach permits studying the structure and dynamics of molecular systems containing millions of atoms \cite{Sanbonmatsu2007, Zhao2013} and evolving on time scales up to hundreds of nanoseconds \cite{Husen_JACS_2016, Salomon2013, Pan2016}. In the molecular mechanics approach, the molecular system is treated classically, i.e. the atoms of the system interact with each other through a parametric phenomenological potential that relies on the network of chemical bonds in the system. This network defines the so-called molecular topology, i.e. a set of rules that impose constraints on the system and permit maintaining its natural shape, as well as its mechanical and thermodynamical properties. The molecular mechanics method has been widely used throughout the past decades and has been implemented, for instance, in the well-established computational packages CHARMM \cite{CHARMMProgram}, AMBER \cite{AMBERprogram}, GROMACS \cite{GROMACSprogram}, NAMD \cite{NAMD} and \MBNExplorer \cite{Solovyov2012}.
\end{sloppypar}

Despite the numerous advantages, standard classical MD cannot simulate irradiation-driven processes. It typically does not account for the coupling of the system to incident radiation, nor does it describe quantum transformations in the molecular system induced by the irradiation. These deficiencies have been overcome recently by introducing Irradiation Driven Molecular Dynamics (IDMD) \cite{Sushko_IS_AS_FEBID_2016}, a new methodology allowing atomistic simulation of IDC in complex molecular systems.
The IDMD approach has been implemented in \MBNExplorer \cite{Solovyov2012}, the advanced software package for multiscale simulations of complex biomolecular, nano- and mesoscopic systems \cite{Solov'yov2017, Solo2017, Solov2017}.

The IDMD methodology is applicable to any molecular system exposed to radiation. It accounts for the major dissociative transformations of irradiated molecular systems (topological changes, redistribution of atomic partial charges, atomic valences, bond multiplicities, interatomic interactions) and possible paths of their further reactive transformations \cite{Sushko_IS_AS_FEBID_2016}. Such transformations can be simulated by means of MD with reactive force fields, particularly with the reactive CHARMM (rCHARMM) force field \cite{Sushko_2016_EPJD.70.12} implemented in \MBNExplorer.

\begin{sloppypar}
This paper provides an overview of the IDMD methodology exploiting the rCHARMM force field and complements it with several illustrative examples published previously \cite{Verkhovtsev_2017_EPJD.71.212, deVera_2018_EPJD.72.147, Sushko_IS_AS_FEBID_2016}.
\end{sloppypar}

\section{Irradiation driven molecular dynamics}
\label{sec:IDMD}

\begin{sloppypar}
This section describes the key principles of the IDMD methodology.
Within the framework of IDMD various quantum collision processes (e.g. ionization, electronic excitation, bond dissociation via electron attachment, or charge transfer) are treated as random, fast and local transformations incorporated into the classical MD framework in a stochastic manner with the probabilities elaborated on the basis of quantum mechanics. This can be achieved because the aforementioned quantum processes happen on the sub-femto- to femtosecond time scales (i.e. during the periods comparable or smaller than a typical single time step of MD simulations) and typically involve a relatively small number of atoms.
\end{sloppypar}

The probability of each quantum process is equal to the product of the process cross section and the flux density of incident particles \cite{LL3}.
The cross sections of collision processes can be obtained from (i) ab initio calculations performed by means of various dedicated codes, or (ii) analytical estimates and models, (iii) experiments, as well as (iv) atomic and molecular databases.
The flux densities of incident particles are usually specific for the problem and the system considered. The properties of atoms or molecules (energy, momentum, charge, valence, interaction potentials with other atoms in the system, etc.) involved in such quantum transformations are changed according to their final quantum states in the corresponding quantum processes.

\begin{sloppypar}
In the course of a quantum process the energy transferred to the system through irradiation is absorbed by the involved electronic and ionic degrees of freedom, and chemically reactive sites (atoms, molecules, molecular sites) in the irradiated system are created.
The follow-up dynamics of the reactive sites may be described by the classical MD and the thermodynamic state of the system until the system undergoes further irradiation-driven quantum transformations. The chemically reactive sites may also be involved in the chemical reactions leading to the change of their molecular and reactive properties, and participate in forming stable and chemically neutral atoms and molecules.
\end{sloppypar}

IDMD simulations permit to account for the dynamics of secondary electrons and the mechanisms of energy and momentum transfer from the excited electronic subsystem to the system's vibrational degrees of freedom, i.e. to its heating. For small molecular systems being in the gas phase the ejected electrons can often be uncoupled from the system and excluded from the analysis of the system's post-irradiation dynamics. For the extended molecular and condensed phase systems, the interaction of secondary electrons with the system can be treated within various electron transport theories, such as diffusion \cite{Surdutovich_2014_EPJD.68.353, Surdutovich_2015_EPJD.69.193} or Monte Carlo (MC) approach \cite{Dapor_2020_MC}, and be considered as additional irradiation field imposed on the molecular system \cite{DeVera2020}. Such an analysis provides the spatial distribution of the energy transferred to the medium through irradiation. Finally, immobilized electrons and electronic excitations transfer the deposited energy to the system's heat via the electron-phonon coupling, which typically lasts up to the picosecond time scale \cite{Gerchikov_2000_JPB.33.4905}. The IDMD approach accounts for the key outcomes of this relaxation process, determining its duration, as well as the temporal and spatial dependence of the amount of energy transferred into the system's heat. As such, IDMD allows the computational analysis of physicochemical processes occurring in the systems coupled to radiation on time and spatial scales far beyond the limits of quantum mechanics-based computational schemes (e.g. DFT and TDDFT, nonadiabatic MD, Ehrenfest dynamics, etc.). This analysis is based on the atomistic approach as any other form of traditional MD.

\begin{sloppypar}
IDMD relies on several input parameters such as the bond dissociation energies, molecular fragmentation cross sections, amount of energy transferred to the system upon irradiation, energy relaxation rate, and spatial region wherein the energy is relaxed. These characteristics, originating from smaller spatial and temporal scales, can be obtained by accurate quantum-mechanical calculations by means of the aforementioned computational schemes. If such calculations become too expensive, the required parameters can still be obtained from experimental data or by means of analytical models/methods.
A similar methodology is also utilized successfully in transport particle models based on the Monte Carlo approach, where the cross sections of various quantum processes are utilized as input parameters for numerous codes simulating particles' transport in various materials.
\end{sloppypar}

Due to the limited number of parameters of IDMD and the reactive molecular force fields, and a much larger number of various output characteristics accessible for simulations and analysis, the IDMD approach opens unique possibilities for modeling of irradiation-driven modifications and chemistry of complex molecular systems beyond the capabilities of either pure quantum or pure classical MD.
By linking outputs of numerous MC codes simulating radiation and particle transport in different media (e.g. Geant4 \cite{Agostinelli2003_Geant4, Allison2016_Geant4}, SEED \cite{Azzolini_2019_SEED}, and others) with the inputs of IDMD one can achieve the multiscale description of irradiation driven chemistry and structure formation in many different Meso-Bio-Nano systems. These important capabilities of \MBNExplorer have been demonstrated in the recent work on the FEBID case study \cite{DeVera2020}. A similar methodology can be used for simulating numerous molecular systems placed into radiation fields of different modalities, geometries and temporal profiles.

\begin{sloppypar}
The developed IDMD framework provides a broad range of possibilities for multiscale modeling of the IDC processes that underpin emerging technologies ranging from controllable fabrication of nanostructures with nanometer resolution \cite{Huth2012, Utke2008, Banqiu2007, hawryluk1988} to radiotherapy cancer treatment \cite{schardt2010heavy, Surdutovich_2014_EPJD.68.353, solov2016nanoscale}, both discussed further in this paper.
\end{sloppypar}

The IDMD algorithm has been validated through a number of case studies of collision and radiation processes including atomistic simulations of the FEBID process and related IDC \cite{Sushko_IS_AS_FEBID_2016, DeVera2020, Prosvetov2021_PtPF3_4}, collision-induced multifragmentation of fullerenes \cite{Verkhovtsev_2017_EPJD.71.212}, electron impact induced fragmentation of organometallic molecule W(CO)$_6$ \cite{deVera_2019_EPJD.73.215}, thermal splitting of water \cite{Sushko_2016_EPJD.70.12}, radiation chemistry of water in the vicinity of ion tracks \cite{deVera_2018_EPJD.72.147}, DNA damage of various complexity induced by ions \cite{Friis2020}, and other \cite{Solov'yov2017}. Several case studies from the aforementioned list are discussed in greater details below in Sections~\ref{sec:C60_fragmentation}--\ref{sec:IDMD_FEBID}.

\section{Reactive CHARMM force field}
\label{sec:rCHARMM}

The reactive CHARMM (rCHARMM) force field \cite{Sushko_2016_EPJD.70.12} implemented in \MBNExplorer enables the description of bond rupture events and the formation of new bonds by chemically active atoms in the system, monitoring all the changes of the system's topology that occur during its transformations.
Being an extension of the commonly used standard CHARMM force field \cite{CHARMM, CHARMM27, CHARMM36}, rCHARMM is directly applicable to organic and biomolecular systems. Its combination with other force fields \cite{Verkhovtsev_2017_EPJD.71.212, Prosvetov2021_PtPF3_4} enables simulations of an even broader variety of molecular systems experiencing chemical transformations whilst monitoring their molecular composition and topology changes \cite{Friis2020, deVera_2018_EPJD.72.147, ES_AVS_2019_EPJD.73.241, Verkhovtsev_2017_EPJD.71.212, Sushko_IS_AS_FEBID_2016, deVera_2019_EPJD.73.215, DeVera2020}.

\begin{sloppypar}
Compared to the standard CHARMM force field, rCHARMM requires the specification of two additional parameters for the bonded interactions, which define the dissociation energy of a covalent bond and the cutoff radius for bond breaking or formation.
By specifying these parameters, \MBNExplorer considers all molecular mechanics interactions (i.e. bonded, angular and dihedral interactions) using an alternative parametrization. If the distance between a given pair of atoms becomes greater than the specified cutoff radius, this particular bonded interaction is removed from the system's topology and not considered in future calculations.
\end{sloppypar}

The standard CHARMM force field \cite{CHARMM} employs harmonic approximation for describing the interatomic interactions, thereby limiting its applicability to small deformations of the molecular system. In case of larger perturbations the potential should decrease to zero as the valence bonds rupture. To permit rupture of covalent bonds in the molecular mechanics force field, \MBNExplorer uses a modified interaction potential for atoms connected by chemical bonds. The standard CHARMM force field describing covalent bonds is defined as
\begin{equation}
U^{({\rm bond})}(r_{ij}) = k_{ij}^{b} (r_{ij} - r_0)^2 \ ,
\label{Eq:CHARMM_Ubond}
\end{equation}
where $k_{ij}^{b}$ is the force constant of the bond stretching, $r_{ij}$ is the distance between atoms $i$ and $j$, and $r_0$ is the equilibrium bond length. This parametrization describes well the bond stretching regime in the case of small deviations from $r_0$ but gives an erroneous result for larger distortions.

\begin{figure}[t!]
\centering
\includegraphics[width=0.45\textwidth]{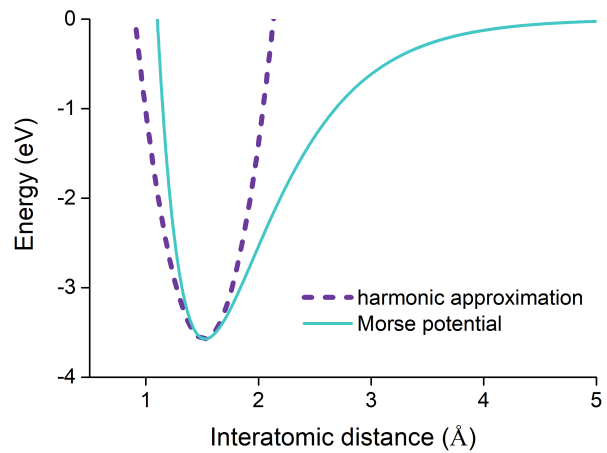}
\caption{The pairwise carbon--carbon (type CN7--CN8B) interaction potential in harmonic approximation~(\ref{Eq:CHARMM_Ubond}) and modeled with the Morse potential~(\ref{Eq:rCHARMM_Morse}).}
\label{fig:rCHARMM_Morse}
\end{figure}

\begin{figure}[t!]
\centering
\includegraphics[width=0.48\textwidth]{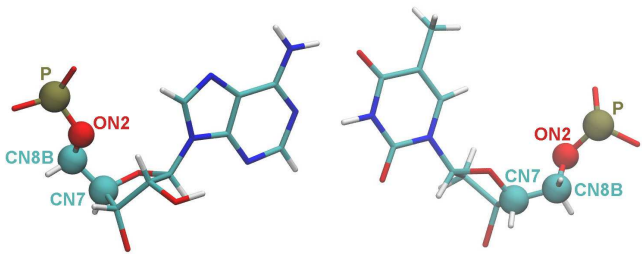}
\caption{Structure of a complementary adenine--thymine nucleobase pair representing a segment of the DNA molecule. Solid spheres highlight the atoms of the DNA backbone with the CHARMM atom type CN7, CN8B, ON2 and P. The bonded, angular and dihedral interaction potentials involving these atoms are shown in Figures~\ref{fig:rCHARMM_Morse} and~\ref{fig:rCHARMM_ang-dih}.}
\label{fig:DNA_nucleobase_pair}
\end{figure}

For a satisfactory description of the covalent bond rupture, it is reasonable to use the Morse potential. This potential requires one additional parameter compared to the harmonic potential (\ref{fig:rCHARMM_Morse}), namely the bond dissociation energy. The Morse potential for a pair of atoms reads as:
\begin{equation}
U^{({\rm bond})}(r_{ij}) = D_{ij} \left[ e^{-2\beta_{ij}(r_{ij} - r_0)} - 2e^{-\beta_{ij}(r_{ij} - r_0)} \right] \ ,
\label{Eq:rCHARMM_Morse}
\end{equation}
where $D_{ij}$ is the dissociation energy of the covalent bond and $\beta_{ij} = \left( k_{ij}^{b} / D_{ij} \right)^{1/2}$ determines steepness of the potential. Figure~\ref{fig:rCHARMM_Morse} illustrates the Morse potential for the CN7--CN8B bond that is one of the covalent bonds in the DNA backbone (see the notations in Fig.~\ref{fig:DNA_nucleobase_pair}). For small deviations from $r_0$ the Morse potential and the harmonic approximation are close to each other.

The rupture of covalent bonds in the course of simulation automatically employs an improved potential for valence angles. In the CHARMM force field, the potential associated with the change of a valence angle between bonds with indices $ij$ and $jk$ reads as:
\begin{equation}
U^{({\rm angle})}(\theta_{ijk}) = k^\theta_{ijk} (\theta_{ijk} - \theta_0 )^2  \ ,
\label{Eq:CHARMM_Uangle}
\end{equation}
where $k^\theta_{ijk}$ and $\theta_0$ are parameters of the potential, and $\theta_{ijk}$ is the actual value of the angle formed by the three atoms. This potential grows rapidly with increasing the angle, and it may lead to non-physical results when modeling the covalent bond rupture. To avoid such cases the harmonic potential, Eq.~(\ref{Eq:CHARMM_Uangle}), is substituted in rCHARMM with the following alternative parametrization:
\begin{equation}
U^{({\rm angle})}(\theta_{ijk}) = 2 k^\theta_{ijk}  \left[ 1 - \cos(\theta_{ijk}-\theta_0 )  \right] \ .
\label{Eq:rCHARMM_angle}
\end{equation}
At small variations of the valence angle this parametrization is identical to the harmonic approximation~(\ref{Eq:CHARMM_Uangle}) used in the standard CHARMM force field. For larger values of $\theta_{ijk}$, the new parametrization~(\ref{Eq:rCHARMM_angle}) defines an energy threshold that becomes important for accurate modeling of bond breakage.

The rupture of a covalent bond is accompanied by the rupture of the angular interactions associated with this bond. The effect of bond breakage on the angular potential can be described through a smoothed step function $\sigma(r_{ij})$ defined as
\begin{equation}
\sigma(r_{ij}) = \frac{1}{2} \left[1-\tanh(\beta_{ij}(r_{ij}-r_{ij}^*))  \right] \ ,
\label{Eq:rCHARMM_Rupture_param}
\end{equation}
with $r_{ij}^*=(R^{{\rm vdW}}_{ij}+r_0)/2$.
This function introduces a correction to the angular interaction potential, assuming that the distance between two atoms involved in an angular interaction increases from the equilibrium value $r_0$ up to the van der Waals contact distance $R^{{\rm vdW}}$. Since an angular interaction depends on two bonds connecting the atoms with indices $ij$ and $jk$, the potential energy describing the valence angular interaction that is subject to rupture is parameterized as
\begin{equation}
\tilde{U}^{({\rm angle})}(\theta_{ijk}) = \sigma(r_{ij}) \, \sigma(r_{jk}) \, U^{({\rm angle})}(\theta_{ijk}) \ .
\label{Eq:rCHARMM_angle-2}
\end{equation}

\begin{figure*}[t!]
\centering
\includegraphics[width=0.85\textwidth]{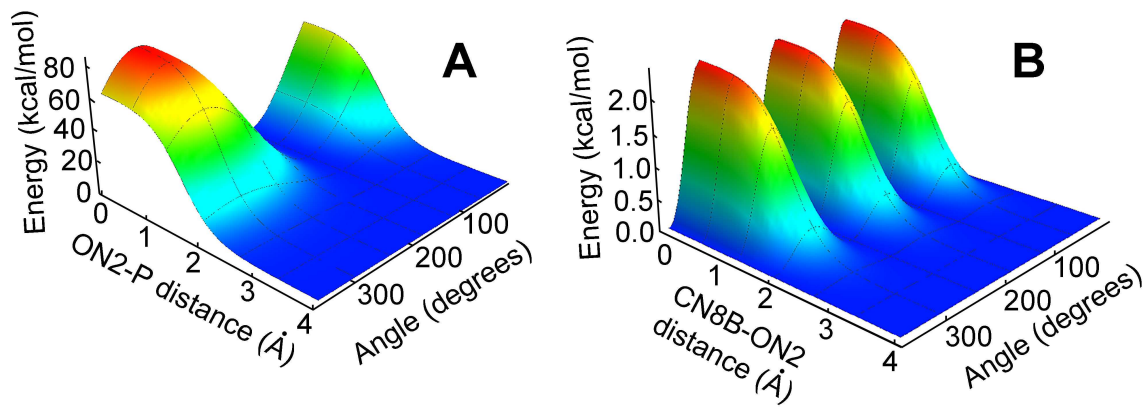}
\caption{\textbf{A:} The CN8B--ON2--P angular potential calculated using Eq.~(\ref{Eq:rCHARMM_angle-2}) with account for the ON2--P bond rupture. \textbf{B:} The CN7--CN8B--ON2--P dihedral potential calculated using Eq.~(\ref{Eq:rCHARMM_dihedral}) with account for the CN8B--ON2 bond rupture. See the atom type notations in Fig.~\ref{fig:DNA_nucleobase_pair}.}
\label{fig:rCHARMM_ang-dih}
\end{figure*}

As seen from Eq.~(\ref{Eq:rCHARMM_angle-2}), the angular potential decreases with the increase of the bond length between any of the two pairs of atoms $ij$ or $jk$. As an illustration, Fig.~\ref{fig:rCHARMM_ang-dih}A shows the CN8B--ON2--P angular potential which arises when modeling the DNA molecule (see the notations in Fig.~\ref{fig:DNA_nucleobase_pair}). The presented angular potential is calculated using Eq.~(\ref{Eq:rCHARMM_angle-2}) assuming the breakage of the bond between the oxygen and the phosphorous atoms. For the sake of illustration, the CN8B--ON2 bond length is taken equal to its equilibrium value $r_0$.

Dihedral interactions arise in the conventional molecular mechanics potential due to the change of the dihedral angles between every four topologically defined atoms. Let us consider a quadruple of atoms with indices $i$, $j$, $k$ and $l$, bound through an interaction governed by a change of the dihedral angle.
In this case, the dihedral angle stands for the angle between the plane formed by the atoms $i$, $j$ and $k$, and the plane formed by the atoms $j$, $k$ and $l$. In the harmonic approximation, the dihedral energy contribution reads as:
\begin{equation}
U^{({\rm dihedral})}_{ijkl} = k_{ijkl}^d  \left[ 1 + \cos{\left( n_{ijkl} \, \chi_{ijkl} - \delta_{ijkl} \right)} \right]  \ ,
\label{Eq:CHARMM_Udihedral}
\end{equation}
where $k_{ijkl}^d$, $n_{ijkl}$ and $\delta_{ijkl}$ are parameters of the potential, and $\chi_{ijkl}$ is the angle between the planes formed by atoms $i$, $j$, $k$ and $j$, $k$, $l$.

\begin{sloppypar}
The dihedral interactions also become disturbed upon covalent bond rupture; therefore, Eq.~(\ref{Eq:CHARMM_Udihedral}) should be modified to account properly for this effect. The rupture of a dihedral interaction between a quadruple of atoms $i$, $j$, $k$ and $l$ should take into account three bonds that contribute to this interaction. Thus, the potential energy describing the dihedral interaction with account for the bond rupture reads as:
\begin{equation}
\tilde{U}^{({\rm dihedral})}_{ijkl} = \sigma(r_{ij}) \, \sigma(r_{jk}) \, \sigma(r_{kl}) \, U^{({\rm dihedral})}_{ijkl} \ ,
\label{Eq:rCHARMM_dihedral}
\end{equation}
where $U^{({\rm dihedral})}_{ijkl}$, given by Eq.~(\ref{Eq:CHARMM_Udihedral}), describes the dihedral interaction within the framework of the standard CHARMM force field. The functions $\sigma(r_{ij})$, $\sigma(r_{jk})$,and $\sigma(r_{kl})$ are defined by Eq.~(\ref{Eq:rCHARMM_Rupture_param}); they are used to limit the dihedral interaction upon increasing the corresponding bond length.
Figure~\ref{fig:rCHARMM_ang-dih}B shows a profile of a CN7--CN8B--ON2--P dihedral interaction potential with accounting for the rupture of the central CN8B--ON2 bond. This dihedral interaction is important for modeling strand breaks in the DNA sugar-phosphate backbone.
\end{sloppypar}

\section{Multi-fragmentation of carbon fullerenes upon collision}
\label{sec:C60_fragmentation}

Irradiation- and collision-induced processes involving carbon fullerenes have been widely studied over the past several decades, both experimentally and theoretically \cite{Campbell_FulCollisions, ISACC_book_2008, Campbell_2000_RepProgPhys.63.1061, Gatchell_2016_JPB.49.162001}.
In particular, collisions involving neutral and charged C$_{60}$ and C$_{70}$ fullerenes have been widely studied \cite{Campbell_FulCollisions, Campbell_1993_PRL.70.263, Rohmund_1996_JPB.29.5143, Brauning_2003_PRL.91.168301, Jakowski_2012_JPCL.3.1536, Handt_2015_EPL.109.63001}. Much experimental information has been obtained in the cited papers on the probability of fullerene fusion and the production of smaller clusters due to subsequent fragmentation.

\begin{sloppypar}
Collision-induced fusion and fragmentation of C$_{60}$ fullerenes was studied in Ref.~\cite{Verkhovtsev_2017_EPJD.71.212} by means of classical MD simulations with the rCHARMM force field, performed with \MBNExplorer.
The many-body Brenner (reactive empirical bond-order, REBO) potential \cite{Brenner90} was used to model interactions between carbon atoms and combined with rCHARMM to monitor changes of the molecular topology and the yields of different atomic and molecular fragments produced.
\end{sloppypar}

Two thousand simulations of $\textrm{C}_{60} + \textrm{C}_{60}$ collisions were performed to reflect the statistical nature of the fusion and fragmentation processes; the duration of the simulation was set to 10~ps.
In each simulation the fullerenes were randomly oriented with respect to each other. The input geometries were prepared by means of \MBNStudio \cite{Sushko2019}, a multi-task software toolkit for \MBNExplorer. The quantitative information on the time evolution of the fragments produced (i.e. the number of fragments of each type) was obtained directly from the output of the simulations. Ensemble-averaged fragment size distribution was calculated for each collision energy by summing up the data from each trajectory and normalizing the resulting value to the total number of fragments.

\begin{figure}[t!]
\centering
\includegraphics[width=0.45\textwidth]{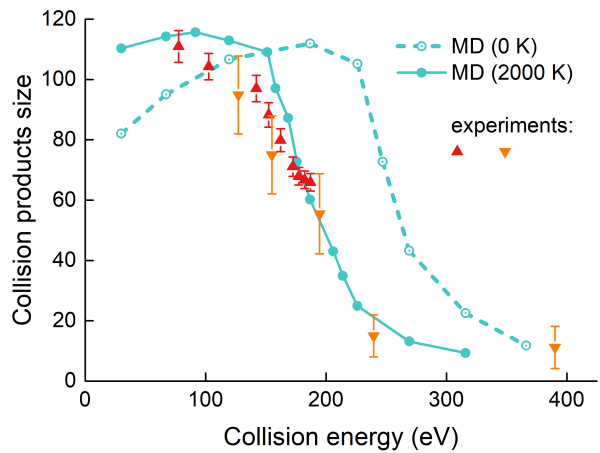}
\caption{The average size of molecular products produced in C$_{60}$-C$_{60}$ collisions as a function of the center-of-mass collision energy. The collision products, including different molecular fragments as well as non-fragmented C$_{60}$ molecules and fused C$_{120}$ compounds, were recorded after 10\,ps of the simulations. Open and filled circles describe the simulations performed at the fullerene initial temperature of 0~K and 2000~K, respectively. Other symbols represent experimental data from Refs.~\cite{Rohmund_1996_JPB.29.5143, Glotov_2000_PRA.62.033202}. In the experiments, an average temperature of the colliding fullerenes was estimated around 2000~K.}
\label{fig_avfragm}
\end{figure}

Figure~\ref{fig_avfragm} shows the average size of the molecular system recorded at the end of the simulations as a function of the center-of-mass collision energy. The average system size was defined as the total number of atoms divided by the total number of molecular species corresponding to given collision energy. Data extracted from the different trajectories at specific collision energy were summed up and normalized to the total number of collision products, including different molecular fragments as well as non-fragmented C$_{60}$ molecules and fused C$_{120}$ compounds.

\begin{sloppypar}
Open circles in Fig.~\ref{fig_avfragm} represent the results obtained at the zero temperature of fullerenes. Figure~\ref{fig_avfragm} shows that the maximal average size of molecular products and hence the maximal fusion probability is obtained at collision energies of about 200\,eV, which is significantly higher than experimental results obtained for ${\textrm C}_{60}^+ + {\textrm C}_{60}$ collisions \cite{Rohmund_1996_JPB.29.5143, Glotov_2000_PRA.62.033202}.
\end{sloppypar}

In the experiments, the average temperature of the colliding fullerenes was estimated around 2000\,K \cite{Rohmund_1996_JPB.29.5143}.
For a better match with the experimental conditions, a set of further simulations was performed where the fullerenes were given an initial temperature of 2000\,K. As a result, each thermally excited molecule had an initial internal kinetic energy of about 30\,eV. Different initial structures and velocities used for the collision simulations were obtained from a 10~ns-long constant-temperature simulation of a single C$_{60}$ molecule being at $T=2000$\,K. Simulations performed at different fullerene temperatures suggest that C$_{60}$ resembles its intact cage-like structure up to $T \approx 2300$~K. At higher temperatures, a transition, which is usually considered as fullerene melting takes place. It corresponds to an opening of the fullerene cage and the formation of a highly distorted but still non-fragmented structure \cite{Kim_1994_PRL.72.2418, Hussien_2010_EPJD.57.207}.

The results of the simulations at $T = 2000$~K are shown in Fig.~\ref{fig_avfragm} by filled circles.
The results reveal that simulations with thermally excited fullerenes agree much better with the experimental results. Taking into account that the statistical uncertainty of the calculated average size of collision products is about 10\%, the calculated numbers agree well with the experimental data. It was found that the largest average product size and hence the highest probability of fusion is for collisions with energies of $90 - 120$~eV, which is significantly lower than the value of about 200~eV simulated at zero initial temperature. The fusion barrier decreases due to the thermal energy stored in the fullerenes.

\begin{figure}[t!]
\centering
\includegraphics[width=0.5\textwidth]{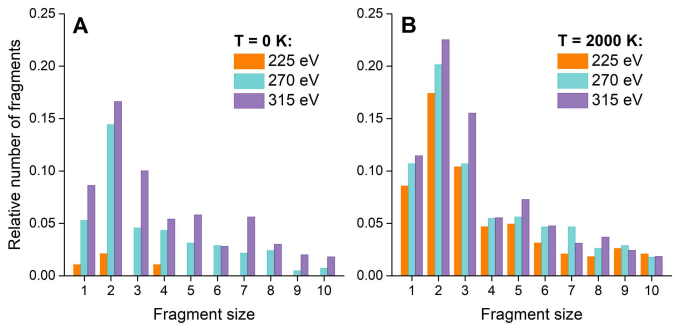}
\caption{Number of C$_n$ ($n \le 10$) fragments, normalized to the total number of fragments produced after 10~ps, for the center-of-mass collision energies of 225, 270 and 315~eV \cite{Verkhovtsev_2017_EPJD.71.212}. Panels A and~B show the results obtained at the 0~K and 2000~K temperature of colliding fullerenes, respectively.}
\label{fig_numberrfrag}
\end{figure}

To analyze the impact of the fullerene initial temperature on the fragmentation dynamics, the size distribution of fragments containing up to 10 carbon atoms formed after 10~ps-long simulations was analyzed. The results of this analysis are shown in Fig.~\ref{fig_numberrfrag}. The simulations where fullerenes had zero temperature before the collision (Fig.~\ref{fig_numberrfrag}A) show that at the collision energy of 225~eV, only a few fragmentation events have been observed. At the energy of 270~eV a phase transition has taken place leading to multi-fragmentation of the fullerenes and the formation of multiple small-size fragments.
The indicated center-of-mass collision energies can be compared with the potential energy difference between the C$_{60}$ fullerene and the gas of 30 C$_2$ dimers, $\Delta E({\rm C}_{60} - 30{\rm C}_2) = 233$~eV, as calculated with the many-body Brenner potential. The collision energy of 225~eV is below the threshold for multifragmentation of C$_{60}$ fullerene. Hence, only a small number of fragments are produced at this energy due to the evaporation process. In contrast, the collision energy of 270~eV exceeds by far the threshold for multifragmentation, thus enabling the formation of many small carbon fragments.
The results of simulations for fullerenes at initial temperature of 2000~K (Fig.~\ref{fig_numberrfrag}B) demonstrate that the phase transition takes place at lower collision energy. In this case the multi-fragmentation regime starts at the collision energy of about 185~eV. The most prominent effect of the fullerene finite temperature is an increase in the number of C$_2$ and C$_3$ fragments. The data shown in Fig.~\ref{fig_numberrfrag} demonstrate that at 315~eV collision energy the relative number of larger fragments is about 3-6\% of the total number of fragments produced, and these values are almost independent of the initial energy stored in the system.

\begin{sloppypar}
It is known that the size distribution of small fragments C$_n$ produced in collisions involving fullerene molecules follows a $n^{-\lambda}$ power law \cite{Rentenier_2005_JPB.38.789, Horvath_2008_PRB.66.075102}. Having taken into account that the simulated distributions of fragments are peaked at $n=2$, the results for $n \ge 2$ were fitted with a power function. The fitting procedure gives the value of $\lambda = 1.47 \pm 0.04$, which is close to the value of 1.54, obtained in earlier MD statistical trajectory simulations at 500~eV center-of-mass collision energy \cite{Schulte_1995_IJMS.145.203}.
\end{sloppypar}

\section{Transport of chemically reactive species around energetic ion tracks}
\label{sec:ion_radiochemistry}

\MBNExplorer can be utilized to evaluate radiobiological damage created by heavy ions propagating in different media, including biological. An energetic charged particle propagating through the medium loses its energy in inelastic collisions with the medium constituents. The transport of produced secondary particles, as well as the radiation damage induced by them are the objects of experimental, theoretical and computational studies \cite{Gustavo_RADAM_book, solov2016nanoscale}. In particular, the physics and chemistry of radiation damage caused by irradiation with protons and heavier ions have recently become a subject of intense interest because of the use of ion beams in cancer therapy \cite{schardt2010heavy, Linz2012_IonBeams, Surdutovich_2014_EPJD.68.353, solov2016nanoscale}.

Radiation damage due to ionizing radiation is initiated by the ions incident on tissue. The initial kinetic energy of the ions ranges from a few to hundreds of MeV per nucleon. In the process of propagation through tissue, they lose energy due to ionization, excitation, nuclear fragmentation, etc. Most of the energy lost by the ion is transferred to the tissue. For an energetic ion propagating in a medium the dependence of the energy deposited into the medium on the penetration distance is characterized by the Bragg peak, that is a sharp maximum in the region close to the end of ion's trajectory. It is commonly understood that the secondary electrons and free radicals produced in the processes of ionization and excitation of the medium with ions are largely responsible for the vast portion of the biodamage. In the Bragg peak region, the secondary electrons lose most of their energy within $1-2$~nm of the ion's path \cite{surdutovich2013biodamage}. After that the electrons continue propagating, elastically scattering with the molecules of the medium until they get bound or solvated electrons are formed \cite{Surdutovich_2014_EPJD.68.353, solov2016nanoscale}. Such low-energy electrons remain important agents for biodamage since they can attach to biomolecules like DNA causing dissociation \cite{Sanche11}.

\begin{figure*}[t!]
\centering
\includegraphics[width=0.85\textwidth]{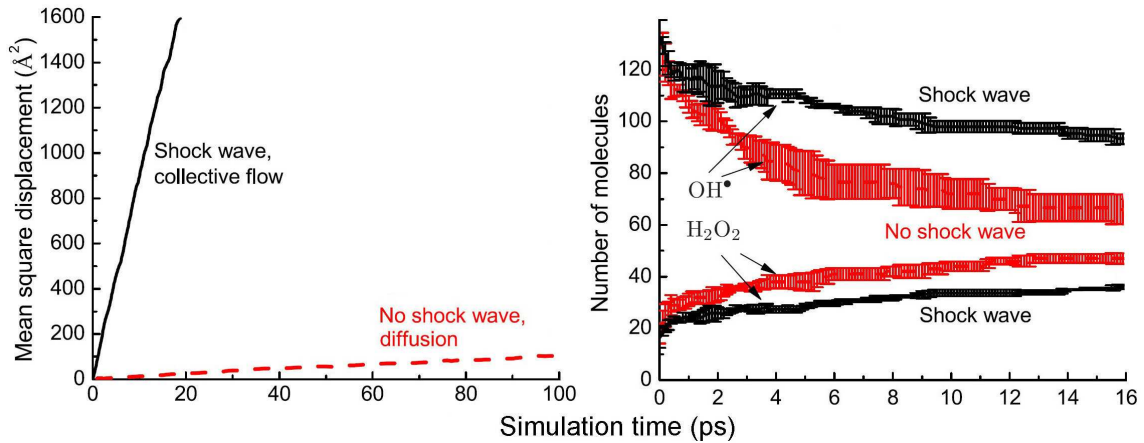}
\caption{\textbf{A:} Mean square displacement of the OH$^{\bullet}$ radicals produced around a 200 keV/u carbon ion path (in the Bragg peak region). Results of simulations where the shock wave is allowed to develop (transport by the collective flow, solid line) and where it is artificially ``switched off'' (transport by diffusion, dashed line) are shown. \textbf{B:} Time evolution of the number of OH$^{\bullet}$ radicals and produced H$_2$O$_2$ molecules. The comparison of the two aforementioned regimes demonstrate the shock wave effects on the radiation chemistry \cite{deVera_2018_EPJD.72.147, de2019role}. }
\label{fig_SW_radiochemistry}
\end{figure*}

The energy lost by secondary electrons in the processes of ionization and excitation of the medium is transferred to its heating (i.e. vibrational excitation of molecules) due to the electron--phonon interaction.
As a result, the medium within the narrow region of the $\sim$1--2~nm radius surrounding the ion's path is heated up rapidly and the pressure within this region increases by several orders of magnitude (e.g. by a factor of 10$^3$ for a carbon ion at the Bragg peak \cite{Toulemonde2009_PRE}) compared to the pressure in the medium outside that region. The pressure builds up by about $10^{-14} - 10^{-13}$~s and it is a source of a cylindrical shock wave \cite{surdutovich2010shock} which propagates through the medium for about $10^{-13} - 10^{-11}$~s.

If the shock wave is strong enough, it can inflict damage by the thermomechanical stress and induce breakage of covalent bonds in the DNA molecule \cite{Yakubovich_2011_AIP.1344.230, surdutovich2013biodamage, devera2016molecular, fraile2019first, bottlander2015effect, Friis2020, Friis2021_SWdamage}. Besides, the radial collective motion of the medium induced by the shock wave is instrumental in propagating the highly reactive molecular species, such as hydroxyl radicals and solvated electrons, to large radial distances (up to tens of nanometers) and slowing down their recombination \cite{Surdutovich_2015_EPJD.69.193, deVera_2018_EPJD.72.147}.

The hydroxyl radical, OH$^{\bullet}$, is one of the main chemically reactive species involved in the indirect damage of DNA molecules. It can form upon fragmentation of ionized (H$_2$O$^+$) or electronically excited water molecules (H$_2$O$^*$):
\begin{eqnarray}
\textrm{H}_2\textrm{O}^+ &\rightarrow& \textrm{OH}^{\bullet} + \textrm{H}^+ \nonumber \\
\textrm{H}_2\textrm{O}^* &\rightarrow& \textrm{OH}^{\bullet} + \textrm{H}^{\bullet}  \ .
\label{water_splitting_reaction}
\end{eqnarray}
The hydroxyl radicals can damage DNA but, among other reactions \cite{hyd2}, they can recombine forming hydrogen peroxide:
\begin{equation}
\textrm{OH}^{\bullet} + \textrm{OH}^{\bullet} \rightarrow \textrm{H}_2\textrm{O}_2 \ .
\label{OH_recombination}
\end{equation}
This reaction was studied earlier~\cite{deVera_2018_EPJD.72.147} by means of the rCHARMM force field implemented in \MBNExplorer. For the sake of simplicity, this reaction was considered as a representative example of the induced radiation chemistry so that all other chemical reactions were disregarded. Pre-solvated electrons and hydrogen radicals were not included in these simulations.

Figure~\ref{fig_SW_radiochemistry}A compares the transport of OH$^{\bullet}$ radicals in the simulations that considered pure diffusion of hydroxyl radicals in the static medium (by artificially switching off the shock wave, see the details in Ref.~\cite{deVera_2018_EPJD.72.147}) and their transport by the collective flow induced by the shock wave. For the latter case, the radicals are transported almost 80 times faster than diffusion, clearly demonstrating the capacity of the collective flows initiated by shock waves to transport effectively reactive species in the medium.

The reactivity of the OH$^{\bullet}$ radicals in the presence of the shock wave is illustrated in Fig.~\ref{fig_SW_radiochemistry}B. It depicts the average number of OH$^{\bullet}$ and H$_2$O$_2$ molecules in the two cases where the shock wave is artificially switched off and naturally allowed to develop, respectively. The error bars represent standard deviations corresponding to three independent simulation runs. As it is seen, the evolution of the number of molecules is different when the collective flow is either present or absent. The transport of radicals by the collective flow does not only propagate the radicals much faster than diffusion but also prevents their recombination, both by their spreading and by creating harsh conditions in which the formation of the O--O bond is suppressed \cite{deVera_2018_EPJD.72.147}. These results are in line with the observations made for the only reported data for water radiolysis with very high-LET ions \cite{LaVerne1996}, where unusually large amounts of radicals escaped ion tracks.

\section{Atomistic simulation of the FEBID process}
\label{sec:IDMD_FEBID}

One of the technological applications of irradiation driven chemistry is focused electron beam induced deposition (FEBID), a novel and actively developing nanofabrication technique that allows the controllable creation of metal nanostructures with nanometer resolution \cite{VanDorp2008, Utke_book_2012, Huth2012}. FEBID is based on the irradiation of precursor molecules (mainly organometallic) \cite{Barth2020_JMaterChemC} by keV electron beams while they are being deposited on a substrate. Electron-induced decomposition of the molecules releases its metallic component, which forms a deposit on the surface with a size similar to that of the incident electron beam (typically a few nanometers) \cite{Utke2008}.

To date, a popular class of precursors for FEBID is metal carbonyls Me$_m$(CO)$_n$ \cite{Huth2012, Kumar2018} which are composed of one or several metal atoms (Me) bound to several carbon monoxide ligands. Metal carbonyls have been widely studied experimentally, and a substantial amount of data on thermal decomposition and electron-induced fragmentation have been accumulated over the past decades \cite{Wysocki_1987_IJMSIP.75.181, Beranova_1994_JAMS.5.1093, Cooks_1990_JASMS.1.16, Wnorowski_2012_IJMS.314.42, Wnorowski_2012_RCMS.26.2093, Neustetter_2016_JCP.145.054301, Lacko_2015_EPJD.69.84}. Strong interest in studying the properties of these compounds has been attributed to their peculiar structure containing strong C--O bonds and relatively weak Me--C bonds. While the former are difficult to cleave, the latter dissociate easily, usually through a sequential loss of CO groups when sufficient internal energy is available.

FEBID operates through successive cycles of precursor molecules replenishment on a substrate and irradiation by a tightly-focused electron beam, which induces the release of metal-free ligands and the growth of metal-enriched nanodeposits.
This process involves a complex interplay of phenomena, taking place on different temporal and spatial scales and thus requiring a dedicated theoretical and computational approach \cite{Solov'yov2017}:
(i) deposition, diffusion, aggregation and desorption of precursor molecules on the substrate;
(ii) transport of the primary, secondary and backscattered electrons;
(iii) electron-induced dissociation of the deposited molecules;
(iv) the follow-up chemistry; and
(v) relaxation of energy deposited into electronic and vibrational degrees of freedom and resulting thermomechanical effects.

\begin{sloppypar}
Until recently, simulations of the FEBID process and the nanostructure growth have been performed by means of the Monte Carlo (MC) approach and the diffusion theory. These methods allow simulations of the average characteristics of the process but do not provide molecular-level details of the adsorbed material. The IDMD approach describes in Sect.~\ref{sec:IDMD} goes beyond this limit and provides a description of the nanostructures created by the FEBID process on the atomistic level \cite{Sushko_IS_AS_FEBID_2016} accounting for the quantum and chemical transformations within the absorbed molecular system.
\end{sloppypar}

In the earlier study \cite{Sushko_IS_AS_FEBID_2016} atomistic IDMD simulations of the FEBID process of tungsten hexacarbonyl W(CO)$_6$ on a hydroxylated SiO$_2$ surface were performed using \MBNExplorer. The results of the simulations were validated through the comparison with experimental data \cite{Fowlkes2010}.

\begin{figure}[t!]
\centering
\includegraphics[width=0.42\textwidth]{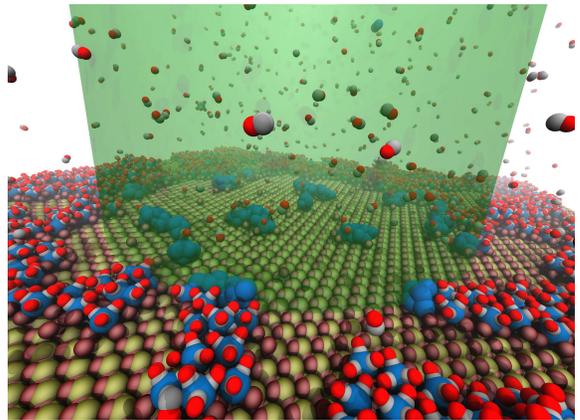}
\caption{Snapshot of the MD simulation \cite{Sushko_IS_AS_FEBID_2016} of adsorption of W(CO)$_6$ precursor molecules atop the SiO$_2$ surface at the early stage of irradiation by an electron beam (a transparent green cylinder). The interaction of deposited precursor molecules with the beam leads to the fragmentation of precursors and to the formation of tungsten clusters, shown in blue. }
\label{fig_FEBID_snapshot}
\end{figure}

The rCHARMM force field was used to model the structure and dynamics of irradiated W(CO)$_6$ molecules atop the SiO$_2$ surface. As describes in Sect.~\ref{sec:rCHARMM}, rCHARMM requires the specification of several parameters, namely the equilibrium bond lengths, bonds stiffness and dissociation energies. Additionally, one needs to define the dissociative chemistry of precursors including the definition of the molecular fragments and atomic valences. In the model considered only the dissociation and formation of the W--C and W--W bonds were permitted, while the C--O bonds were treated within the harmonic approximation, Eq.~(\ref{Eq:CHARMM_Ubond}), preventing those bonds from breakage.

Figure~\ref{fig_FEBID_snapshot} shows a snapshot of the MD simulation of the first irradiation phase, where the irradiation by an electron beam of the cylindrical shape has been considered. Only precursors molecules located inside the cylinder have been exposed to radiation that may induce their dissociation. The dissociation rate of the precursor molecules was evaluated from the experimental data \cite{Fowlkes2010}.

\begin{sloppypar}
The FEBID process was modeled \cite{Sushko_IS_AS_FEBID_2016} with the rescaled computationally accessible parameters (the irradiation time and the beam current). These parameters may differ from experimental values, but they have to be chosen to correspond to a given (in experiment) number of electrons $N_e$ targeting the system (the electron fluence) and thus producing the irradiation induced effects on the same scale as in the experiment. Following this idea, the irradiation time in IDMD simulations is typically decreased as compared to the corresponding experimental values.
\end{sloppypar}

\begin{figure*}[t!]
\centering
\includegraphics[width=0.8\textwidth]{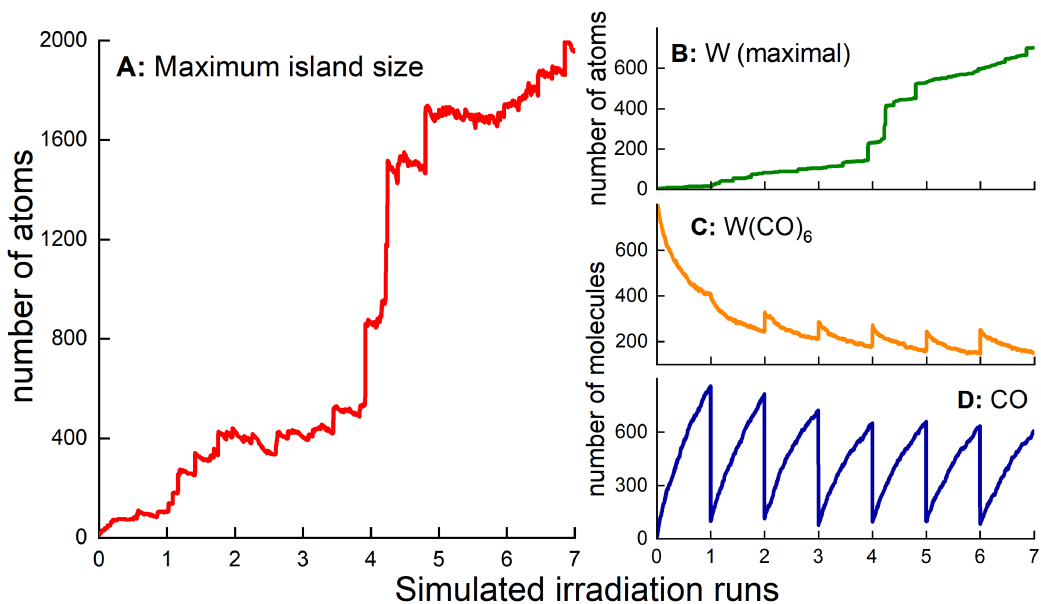}
\caption{Time evolution of the size of the largest W-enriched island \textbf{(A)}, the number of W atoms \textbf{(B)}, the number of W(CO)$_6$ \textbf{(C)} and CO \textbf{(D)} molecules in the system during the irradiation periods \cite{Sushko_IS_AS_FEBID_2016}. The irradiation periods are marked by the successive numbers. The duration of each period is 10~ns. In simulations the electron beam radius $R$ is equal to 5~nm and the beam current $I_0 = 4$~$\mu$A. }
\label{fig_FEBID_island_size}
\end{figure*}

The probability of W--C bonds dissociation in W(CO)$_6$ caused by collisions with electrons per unit time can be evaluated as
\begin{equation}
P = \sigma \times J_0 \ ,
\label{eq:fragm_prob}
\end{equation}
where $\sigma$ is the bond dissociation cross section and
\begin{equation}
J_0 = \frac{I_0}{e \times S_0} \ ,
\label{Eq:FEBID_flux}
\end{equation}
is the electron flux density targeting the precursor. Here $I_0$ is the electron beam current, $e$ is the electron charge and $S_0 = \pi R^2$ is the beam cross section equal with $R$ being the beam radius.
Substituting Eq.~(\ref{Eq:FEBID_flux}) into Eq.~(\ref{eq:fragm_prob}) one derives
\begin{equation}
P_0 = \frac{I_0}{e} \, \frac{\sigma}{S_0} \ .
\end{equation}
Estimating $I_0 = 4$~$\mu$A and $\sigma = 1.2 \times 10^{-2}$~nm$^2$ \cite{Fowlkes2010}, one derives $P_0 = 3.8 \times 10^{-6}$ fs$^{-1}$.
This analysis can be refined on the basis of the accurate calculations of the corresponding fragmentation cross section of the precursor, as well as the yield and the spatial distribution of the secondary electrons, which all depend on the energy of the primary beam \cite{DeVera2020}. The cited study also demonstrated the possibility of linking the input characteristics for the IDMD simulations with outputs of MC codes simulating radiation and particle transport in different media and thus achieving the multiscale description of IDMD of the FEBID process. This methodology is general and can be used to simulate numerous molecular systems placed into radiation fields of different modalities, geometries and temporal profiles.

Figure~\ref{fig_FEBID_island_size}A shows the time dependence of the size of the largest W-enriched island emerged in the simulations \cite{Sushko_IS_AS_FEBID_2016}. Panels B, C and D show this evolution for the numbers of W atoms, W(CO)$_6$ and CO molecules in the system during the irradiation periods, respectively. The irradiation periods are marked in the plots by the consecutive numbers. The duration of each period is 10~ns. The replenishment periods are excluded from the plots. However, the drops in the number of CO molecules and the number of W(CO)$_6$, corresponding to the changes of the system that occur during the replenishment periods, are well seen. Both the size and the number of W atoms in the islands grow due to the attachment of new atoms to the islands during the FEBID process. Figures~\ref{fig_FEBID_island_size}A and ~\ref{fig_FEBID_island_size}B indicate the coalescence of smaller islands into a larger single nanostructure (ripening process) during the initial stage of the FEBID process. The irregular spikes on the curves arise at the instants when separate islands merge.

The W--(CO) bonds in precursors dissociate during the irradiation periods leading to the appearance of the CO molecules. Most of these are created in the vicinity of the surface and evaporated later into the vacuum chamber. The evaporation process continues during the replenishment periods. Figure~\ref{fig_FEBID_island_size}D shows that during each irradiation period the number of CO molecules grows nonlinearly. To account for the evaporation process of CO during the replenishment periods, the CO molecules have been removed from the simulation box after each irradiation phase. After that the new W(CO)$_6$ molecules have been deposited on the surface according to the chosen deposition rate and the duration of the deposition process. This results in the abrupt decrease of the CO molecules and the increase of W(CO)$_6$ numbers before starting each new cycle irradiation.

The results reviewed here demonstrate that the IDMD approach provides a powerful computational tool to model the growth process of W-granular metal structures emerging in the FEBID process at the atomistic level of detail \cite{Sushko_IS_AS_FEBID_2016}. The morphology of the simulated structures, their composition and growth characteristics are consistent with the available experimental data. The performed analysis also indicates the need for further wide exploitation of the IDMD methodology in FEBID and many other processes in which the irradiation of molecular systems and irradiation driven chemistry play the key role.

\section{Conclusions}

Irradiation Driven Molecular Dynamics is a unique computational methodology implemented in the \MBNExplorer software package, enabling atomistic-level modeling of irradiation-driven transformations involving complex molecular systems. The concept of IDMD is general and applicable to any molecular system treated with any classical force field implemented in the software. IDMD is capable of describing systems modeled through pairwise potentials, many-body force fields, molecular mechanics force fields (including the reactive CHARMM force field) and their combinations. The limited number of parameters determining molecular force fields and their irradiation-driven perturbations results in a countable number of modifications that could occur in a molecular system upon irradiation, thus making the method efficient and accurate. This implementation opens a broad range of possibilities for modeling irradiation-driven modifications and chemistry of complex molecular systems.

\section*{Acknowledgements}

\begin{sloppypar}
The authors are grateful for financial support from the COST Action CA17126 ``Towards understanding and modelling intense electronic excitation'' (TUMIEE), Deutsche Forschungsgemeinschaft (Projects no. 415716638, SFB1372 and GRK1885), the Volkswagen Stiftung (Lichtenberg professorship to IAS), and the European Union's Horizon 2020 research and innovation programme -- the Radio-NP project (GA 794733) within the H2020-MSCA-IF-2017 call and the RADON project (GA 872494) within the H2020-MSCA-RISE-2019 call.
\end{sloppypar}


\bibliography{bibliography}

\end{document}